\documentclass[utf8]{FrontiersinHarvard} % for articles in journals using the Harvard Referencing Style (Author-Date), for Frontiers Reference Styles by Journal: https://zendesk.frontiersin.org/hc/en-us/articles/360017860337-Frontiers-Reference-Styles-by-Journal
\usepackage{url,hyperref,lineno,microtype,subcaption}
\usepackage[onehalfspacing]{setspace}
\usepackage{multirow}
\def\keyFont{\fontsize{8}{11}\helveticabold }
\def\firstAuthorLast{Sample {et~al.}} %use et al only if is more than 1 author
\def\Authors{Xiangyu Yan\,$^{1,*}$,  Gongen Han\,$^{1}$, Can Fang\,$^{2}$, and Xuan Jing\,$^{3}$,}
\def\Address{$^{1}$Xi’an GAOXIN High School, Xi’an, China \\
$^{2}$Southwest University, Chongqing, China \\
$^{3}$NCS Pte Ltd, Singapore  }
\def\corrAuthor{Xiangyu Yan}

\def\corrEmail{jason.xyyan@gmail.com}

\begin{document}
\onecolumn
\firstpage{1}

\title {Multi-Interactive-Modality based Modeling for Myopia Pro-Gression of Adolescent Student} 

\author[\firstAuthorLast ]{\Authors} %This field will be automatically populated
\address{} %This field will be automatically populated
\correspondance{} %This field will be automatically populated

\extraAuth{}% If there are more than 1 corresponding author, comment this line and uncomment the next one.
%\extraAuth{corresponding Author2 \\ Laboratory X2, Institute X2, Department X2, Organization X2, Street X2, City X2 , State XX2 (only USA, Canada and Australia), Zip Code2, X2 Country X2, email2@uni2.edu}

\maketitle

\begin{abstract}

%%% Leave the Abstract empty if your article does not require one, please see the Summary Table for full details.
\section{}
Myopia is a common visual disorder that affects millions of people worldwide and its prevalence has been increasing in recent years. Environmental factors, such as reading time, viewing distance, and ambient lighting, have been identified as potential factors in the development of myopia. In this study, we investigated the relationship between three major factors and myopia in 120 adolescents. By collecting environmental images of the adolescents in the learning state as well as retinal fundus images, we proposed an environmental visual load (EVL) model to extract the potential information in these images. Through experimental data analysis, we found that these three major factors are closely related to the severity of myopia, and that the simultaneous exacerbation of these factors sharply increases the myopia of the eye. Our results suggest that interventions targeting these environmental factors may help prevent and manage myopia.

\tiny
 \keyFont{ \section{Keywords:} myopia, reading time, viewing distance, ambient lighting, and environmental visual load model} %All article types: you may provide up to 8 keywords; at least 5 are mandatory.
\end{abstract}

\section{Introduction}

Myopia, or nearsightedness, is a common refractive error that affects millions of people worldwide. In myopia, light entering the eye focuses in front of the retina instead of on it, causing distant objects to appear blurry. While myopia is generally correctable with glasses or contact lenses, high levels of myopia increase the risk of serious eye conditions such as retinal detachment, glaucoma, and cataracts\citep{grosvenor2007primary, wang2023short, han2022myopia}.

\begin{figure}[!htbp]
	\center
	\includegraphics[scale=0.53]{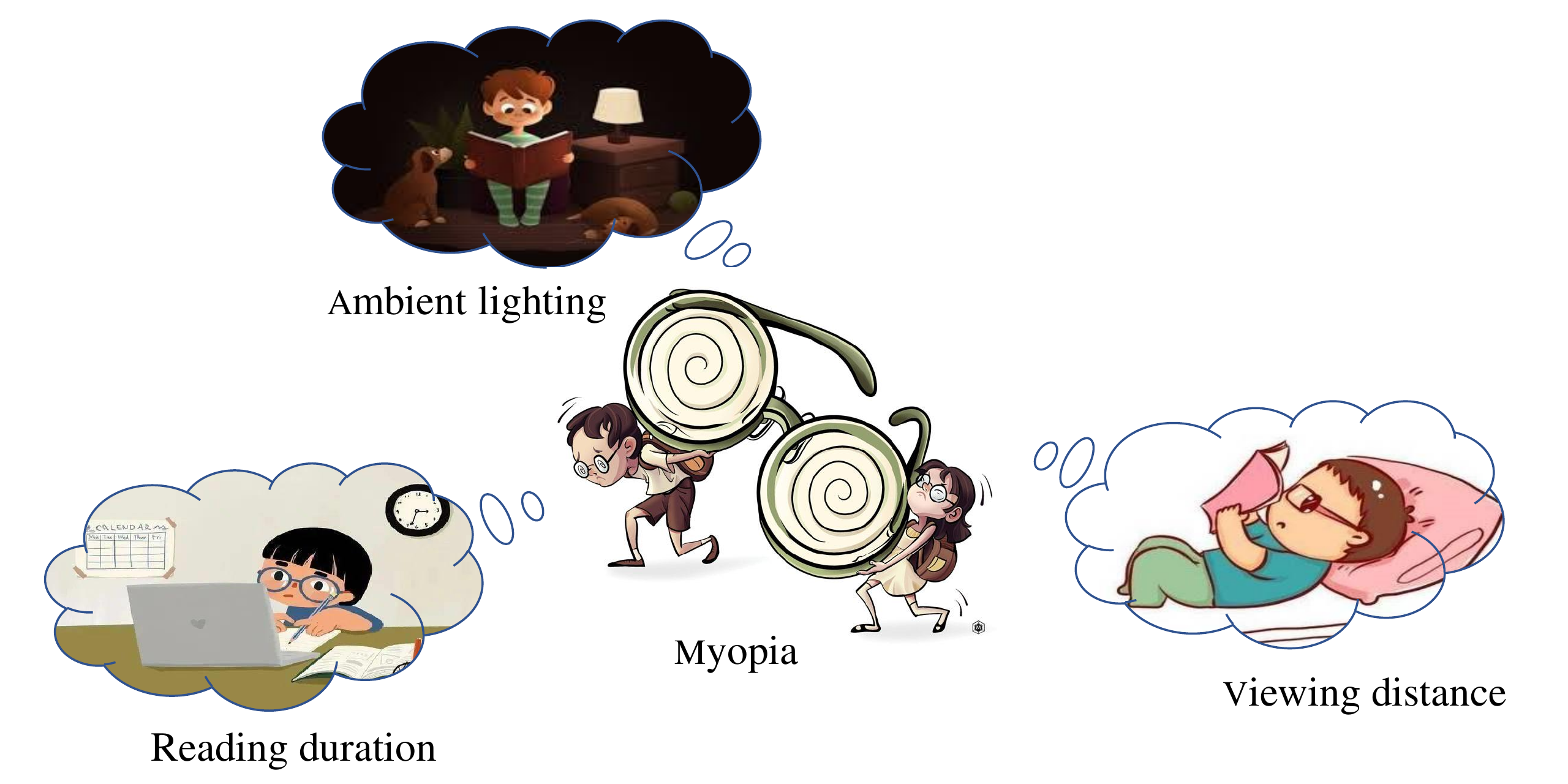}
	\caption{Environmental factors that cause myopia.}
	\label{fig-myopia-factors} 
\end{figure}

The etiology of myopia is multifactorial, with both genetic and environmental factors playing a role. Studies have shown that environmental factors can significantly influence the development and progression of myopia\citep{oner2016influence}. Among these environmental factors are reading duration, viewing distance, and ambient lighting, as shown in Figure \ref{fig-myopia-factors}.

One of the most consistent risk factors for myopia is prolonged near work, such as reading or computer use. This association has been observed in numerous studies across different populations and age groups. The exact mechanisms by which near work contributes to myopia are not fully understood, but it is thought to be related to the accommodative and convergence responses of the eye\citep{singh2019prevalence}. The accommodative response is the process by which the eye adjusts the shape of its lens to focus on near objects. The convergence response is the process by which the eyes rotate inwards to maintain single binocular vision \citep{huang2020protective}. Therefore, the accommodative and convergence responses of the eyes are thought to play a role in the development of myopia due to prolonged near work activities.

The distance between the eyes and the object being viewed is another factor that has been linked to myopia. Studies have found that people who hold reading material closer to their eyes are more likely to be myopic than those who hold it further away \citep{pan2018types}. This relationship may also be related to the accommodative and convergence responses of the eye. The closer an object is to the eyes, 
the greater the demand on the accommodative and convergence systems \citep{li2016effects,jiang2002models}. This increased demand may cause 
the responses to become unbalanced, leading to axial elongation of the eye and the development of myopia. as shown in Fig. 2.

\begin{figure}[!htbp]
	\center
	\includegraphics[scale=0.5]{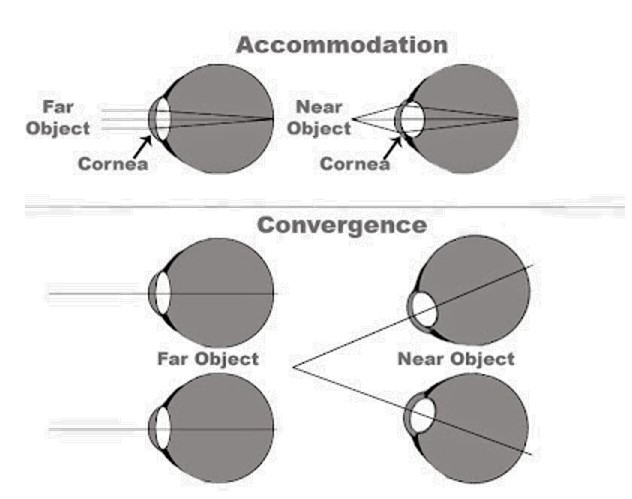}
	\caption{ Example of accommodative and convergence reactions.}
	\label{fig-acc-con} 
\end{figure}

Ambient lighting, or the level of light in the environment, has also been implicated in the development and progression of myopia. Studies have found that people who spend more time in low light environments, such as dark classrooms or  offices, are more likely to be myopic than those who spend more time in brightly lit environments. The exact mechanisms by which ambient lighting affects myopia are not fully understood, but it is thought to be related to the size of the pupil and the amount of aberrations in the eye. In low light environments, the pupil may dilate to let in more light, which can increase the number of aberrations in the eye and contribute to the development of myopia \citep{li2015near, smith2012protective}.

In summary, reading duration, viewing distance, and ambient lighting are all factors that have been implicated 
in the development and progression of myopia \citep{vienne2014effect,matsuo1992follow,borsting2003association}. However, explanations for most of these factors are separate, unsystematic, and lack mathematical modeling. 
In order to gain more insight into the relationship between myopia and these three factors, in this paper, we systematically model the association between myopia and the three factors and propose an environmental visual load (EVL) model. Specifically, we first collected environmental images of adolescents in the learning state as well as retinal fundus images. Second, we analyzed the values of several key variables affecting myopia in terms of reading time, viewing distance, and environmental lighting. Furthermore, based on these variables as well as the image data, we proposed three different models, namely the integrated dual-focus model, the expanded hyperbolic model, and the lighting model, to explore the mathematical relationships between these factors and myopia. In order to be able to describe these relationships more intuitively, we further unify these three models and proposed the final EVL model.

The experimental results showed that reading time, viewing distance, and environmental lighting are all important environmental factors that affect the development of myopia. Changes in these three major factors can lead to a tendency for myopia to develop in children's vision. Our proposed environmental visual load model can counteract this trend by multi-factor modeling.

\section{Methods}

\begin{figure}[!htbp]
	\center
	\includegraphics[scale=0.7]{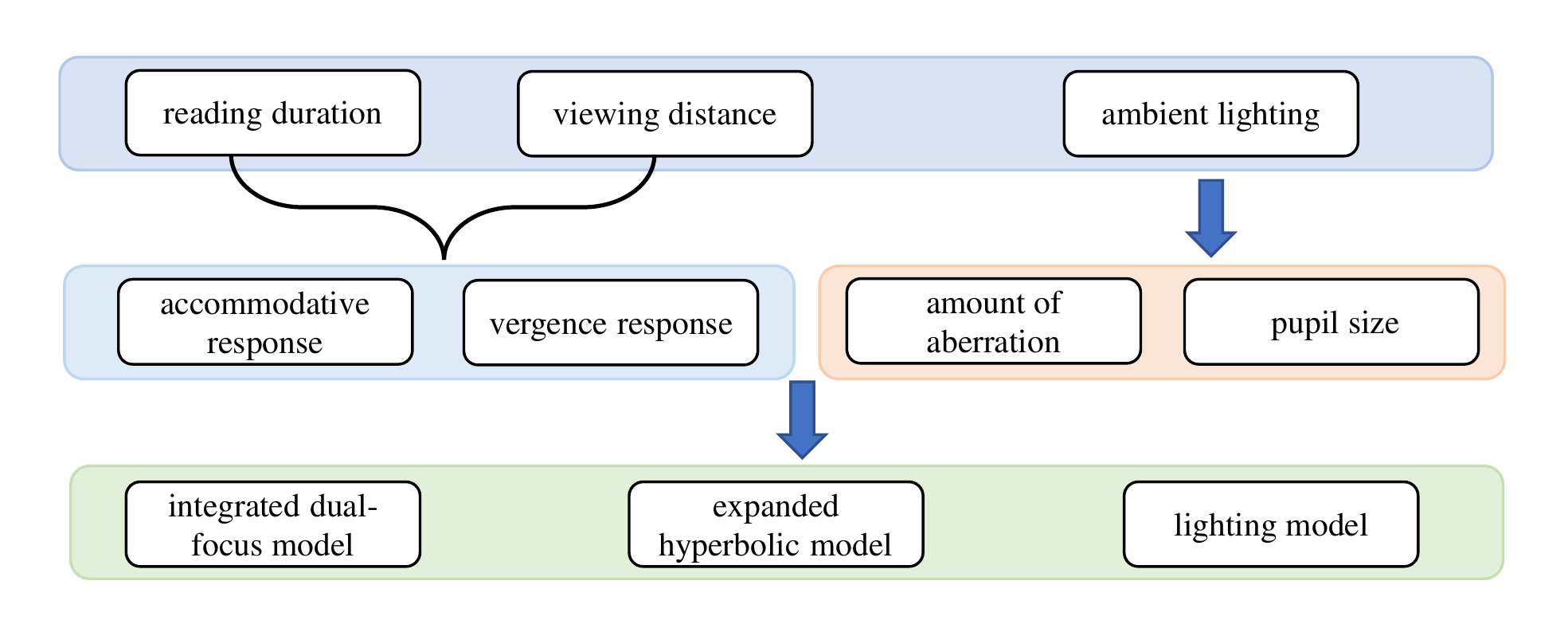}
	\caption{ The flow of the whole system.}
	\label{fig-flow-pdf} 
\end{figure}

For the whole modeling process, as show in Figure \ref{fig-flow-pdf}, we start with the extraction of key attributes. We consider that when people work in close proximity for long periods of time, it leads to a break in the accommodative and convergence response of the eye, and it has been found that children who had a greater accommodative response than convergence response were more likely to develop myopia. Both are therefore included in the overall set of key attributes. As for the effect of ambient light on myopia, we analyzed that the pupil may dilate in low-light environments to let in more light, which may increase the aberration of the eye and lead to the development of myopia. Therefore, we included the amount of aberration and pupil size as key elements.

After obtaining some key variables about myopia, we next model them to establish the intrinsic connections that have responded to the trends in visual acuity in adolescents. The integrated dual-focus model (IDFM) is the first mathematical model we propose that considers the accommodation and convergence responses of the eye during near work. This model assumes that the eye has a resting point of accommodation (RPA) and a resting point of convergence (RPV), which are the positions of the eye's lens and eye muscles, respectively, when the eye is at rest.

The model predicts the resting point of accommodation of the eye, $A$, and the resting point of convergence, $V$ , based on the viewing distance, $d$, and the refractive error of the eye, $M$. The equations for the resting point of accommodation and convergence are as follows:

\begin{equation}
 \begin{aligned}
	& A =  \frac{M}{1-d}\\
	& V =  \frac{M}{1+d}\\
	\label{eq_idfm}
 \end{aligned}
\end{equation}

The IDFM model suggests that when the demand on the accommodative and convergence systems is unbalanced, the eye will respond by elongating axially, which can lead to the development of myopia. Specifically, the model predicts that myopia will develop when the RPA is closer to the eye than the RPV, causing the accommodative response to be greater than the convergence response.

The expanded hyperbolic model (EHM) is another mathematical model we propose that considers the accommodative and convergence responses of the eye, as well as the effects of near work on axial elongation. This model assumes that the elongation of the eye is proportional to the resting point of accommodation, $A$, and inversely proportional to the resting point of convergence, $V$.

The model predicts the axial length of the eye, $AL$ , based on the resting point of accommodation and convergence, the initial axial length of the eye,  $AL_0$, and the duration of near work, $t$. The equation for the axial length of the eye is as follows:

\begin{equation}
	\begin{aligned}
		& AL = AL_0 + n\times t \times  \frac{A}{V}  \\
		\label{eq_ehm}
	\end{aligned}
\end{equation}

\noindent where  $n$  is a constant that depends on the individual and the type of near work being performed.

The EHM model suggests that myopia will develop when the demand on the accommodative and convergence systems is unbalanced, leading to increased axial elongation of the eye. 

The lighting model is a mathematical model that considers the effects of ambient lighting on the size of the pupil and the number of aberrations in the eye. This model assumes that the number of aberrations in the eye, $W$ , is proportional to the size of the pupil, $P$.

The model predicts the refractive error of the eye, $M$ , based on the number of aberrations in the eye,  $W$, the size of the pupil, $P$, and the level of ambient lighting,  $L$. The equation for the refractive error of the eye is as follows:

\begin{equation}
	\begin{aligned}
		& M = M_0 + n \times \frac{(W-W_0) \times (P-P_0)}{L}  \\
		\label{eq_ehm}
	\end{aligned}
\end{equation}

\noindent where  $M_0$ is the initial refractive error of the eye,  $P_0$ is the initial size of the pupil, $W_0$ is the initial number of aberrations,  and  n is a constant that depends on the individual and the type of near work being performed.

The lighting model suggests that myopia will develop when the level of ambient lighting is low, leading to increased pupil size and aberrations in the eye.

The above three equations can model each variable affecting myopia. Next, we will unify the three models into an environmental visual load (EVL) model.

In the previous analysis, we have illustrated two of the three main environmental factors affecting visual acuity, namely reading duration, and viewing distance, both lead to an imbalance in the demands of the accommodative and convergence systems. Therefore, we believe that such an imbalance can be a direct response to the cause of myopia formation, and we also incorporated the effect of ambient light. The EVL model can be calculated as follows:

\begin{equation}
	\begin{aligned}
		& O = \frac{AR}{VR}  \\
		\label{eq_ehm}
	\end{aligned}
\end{equation}

\noindent where $AR$ denotes the accommodative response, and $VR$ denotes the convergence response, and their values change according to the environment of the human eye, so we use the axial length of the eye, $AL$, and the viewing distance, $d$, to correspond to their changes. They are calculated as follows:

\begin{equation}
	\begin{aligned}
		& AR =  AL + M(1-d) \\
		& VR =  AL + M(1+d) \\
		\label{eq_ehm}
	\end{aligned}
\end{equation}

If $ O > 1+ \theta$ or  $  O < 1 - \theta$ , we consider that there is an imbalance between the accommodative and convergence responses, resulting in a reduced accommodative response relative to the level of stimulus convergence. This means that the eyes have difficulty focusing on distant objects, leading to blurry vision. If $  1 - \theta \leq O \leq 1+ \theta$ , then we believe that there is a balance between the adaptive and convergence responses and that there is not yet a clear tendency to cause blurred vision. We set the value of  $\theta$  to 0.1

In order to be able to evaluate the above proposed model accurately, we divided the parameters involved in the model into two categories: computational quantities as well as statistical quantities, as shown in Table \ref{table-params}. The former is based on the calculation of the statistics, while the latter is the value obtained through experimental statistics in this paper. Therefore, in the subsequent experiments, we mainly focus on these statistics for the analysis.

\section{Experiment}

\begin{figure}[!htbp]
	\center
	\includegraphics[scale=0.7]{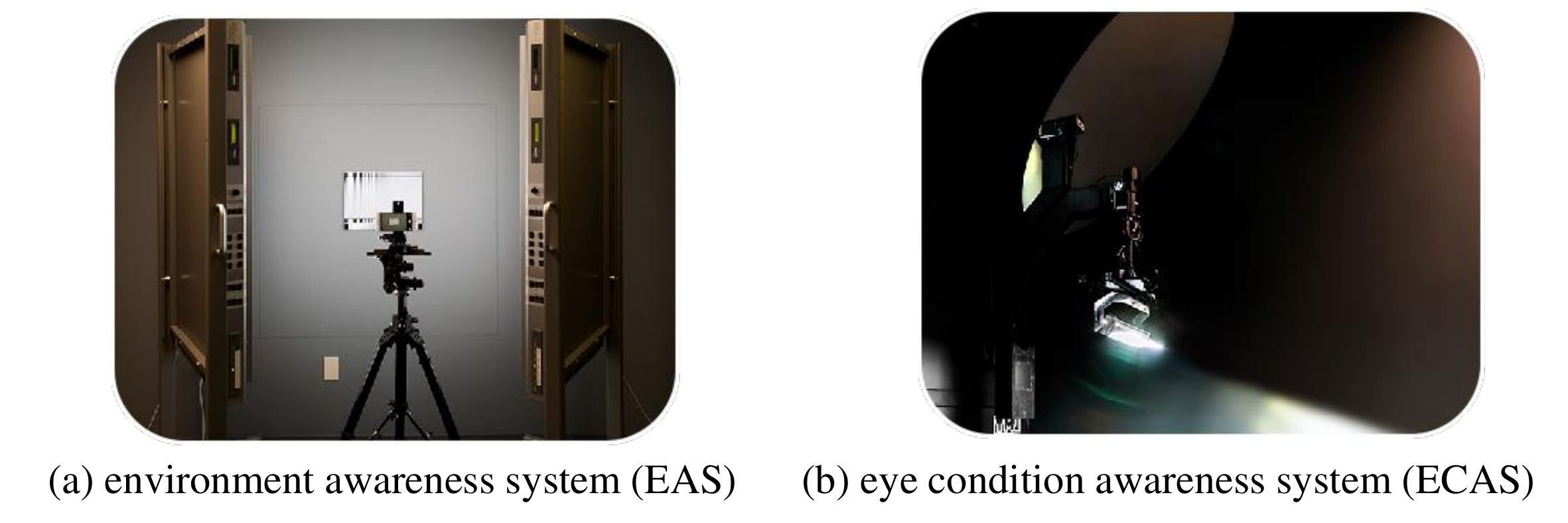}
	\caption{ Experimental Scenario.}
	\label{fig-EAS_ECAS} 
\end{figure}

The experimental scenario is shown in Figure  \ref{fig-EAS_ECAS}. The experimental environment was set up as two scenarios: the Environmental Awareness System (EAS) and the Eye Condition Awareness System (ECAS). The former was mainly used to record data on the subject's environmental state, as well as the ambient light level to which the subject was exposed, and the duration of near work. The latter was used to photograph and measure the subject's eye condition and to obtain retinal fundus images, as well as to analyze parameters such as eye pupil size, refractive error of the eye, aberration size and viewing distance. As for the types of near work of the subjects being performed, they were divided into three types: reading, writing, and playing with cell phones, with corresponding values of 1, 1.5, and 2.

\begin{figure}[!htbp]
	\center
	\includegraphics[scale=0.7]{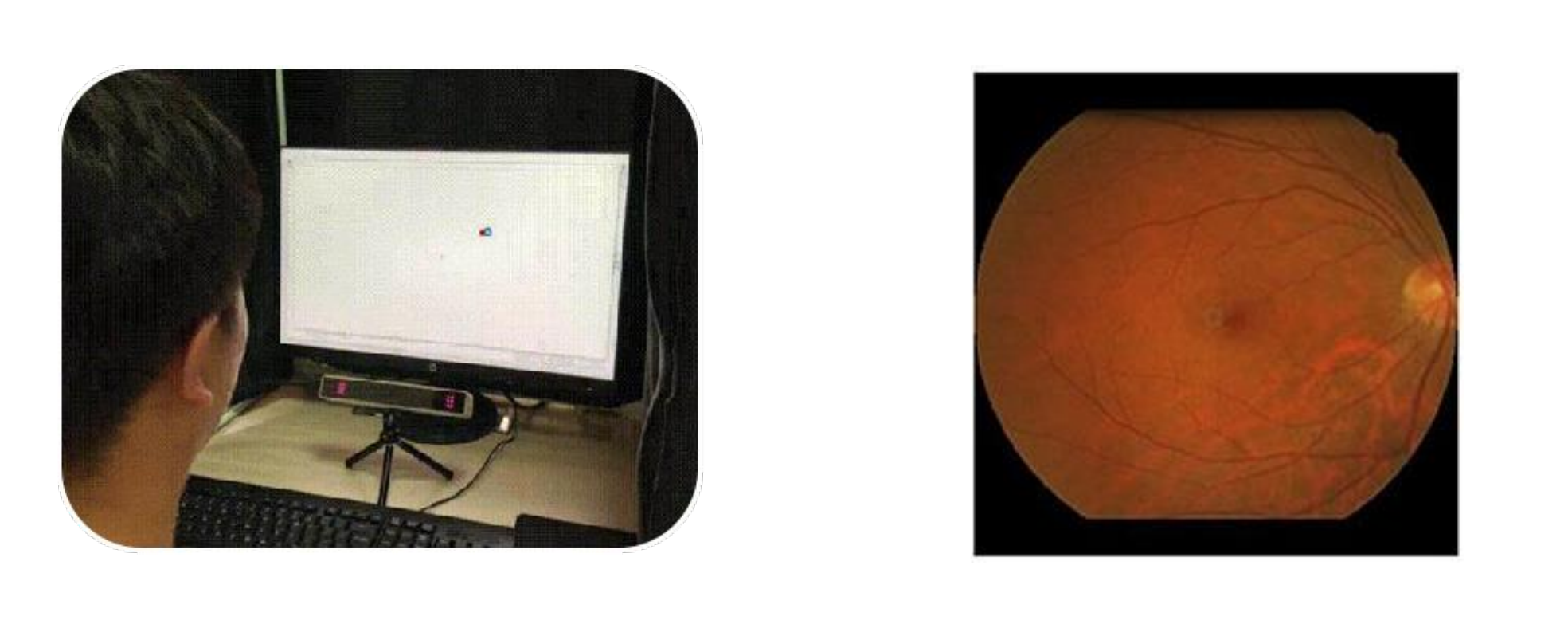}
	\caption{ Example of experiment. The image on the left is an environmental image, and the right is a retinal fundus photograph.}
	\label{fig-example} 
\end{figure}

We conducted a cross-sectional study of 120 adolescents between the ages of 8 and 16, with an average of 15 of each age. Each participant underwent a comprehensive eye examination, including a measurement of spherical equivalent refraction (SER), which is a standard measure of myopia. We also use the EAS system and ECAS system to collect information on the ocular status of each participant in the experimental setting and fill in the corresponding statistics, respectively, as shown in Figure \ref{fig-example} and Table \ref{table-ablation_1}.
A sample of our experimental environment configuration and the retinal fundus photograph collected are shown in Figure \ref{fig-example}.

We divided the collected data into different age groups, with approximately 20 individuals in each age group. And we placed the adolescents of each age group in different experimental settings. Since they differ in their own eye status, they differ in the initialization values of the model. We set three different values for the initial level of ambient lighting $L_0$, namely 189, 527 and 892. Table \ref{table-ablation_1} below shows the experimental initialization values for different age groups of adolescents. 
The data in the table shows the initial axial length of the eye, the size of the pupil, the refractive error of the eye and the amount of aberration, all of which change with age. Axial length usually increases as the eye grows and develops, while the size of the pupil decreases with age, as has been verified in a number of studies \citep{bach2019axial, KASTHURIRANGAN20061393}.

The experimental data collected through the experimental system are shown in Table \ref{table-ablation_2}. We still divided the collected data into different age groups. And for the colected data, we divided them according to different types of experimental systems, e.g., the EAS system mainly collected the duration of near work $t$ and the level of ambient lighting $L$, while the ECAS system mainly collected the size of the pupil $P$, viewing distance $d$ , the refractive error of the eye  $M$ and the number of aberrations $W$.

After obtaining all the experimental data, we calculated and analyzed them in detail. In the 8-10 years old group, we mainly conducted experiments on the duration of near work. With the increase of work time, the accommodative and convergent responses were unbalanced, and the response was based on the $O$ value, and the SER value showed a downward trend. This indicates that our EVL model can reflect the adverse effects of the duration of near work on visual acuity.

% Our results showed a significant negative correlation between the duration of near work and myopia, with an imbalance between the accommodative and convergent responses when the duration of near work increased in the 8-10-year-old group, a change that was consistent with the trend of decreasing SER values.

This phenomenon was also present in the ambient lighting and reading distance factors, which were manifested in the 10-12 and 12-14-year-old groups, respectively. The absolute value of the trend toward myopia gradually increased as ambient light became weaker and reading distance became shorter. And in the 14-16 years old group, we found that increasing the values of the three main factors simultaneously led to an increase in the output values of the final EVL model, which increased the imbalance between the adaptive and convergent responses. All these groups led to a decrease in the value of SER indicators with the change of factors. And the final 14-16 group showed the most significant changes, also indicating that simultaneous changes in reading time, ambient lighting and reading distance had the greatest effect on children's visual acuity.

\section{CONCLUSIONS}
Our findings suggest that reading time, viewing distance, and environmental lighting are all important environmental factors that influence the development of myopia. The development of myopia. With these three main factors in mind, we analyzed the key ocular attributes that would be affected by these three factors. By introducing the EAS and ECAS devices, we were able to analyze these key attributes using the acquired environmental images as well as retinal fundus images. Based on this we proposed three different models: integrated dual-focus model, expanded hyperbolic model, and lighting model, which were combined into the final environmental visual load (EVL) model. The EVL model was used to reflect the trend of myopia.

\begin{table}[!htbp]
	\center
	\caption{Parameters of each model}
	\resizebox{0.7\linewidth}{!}{
		\begin{tabular}{l|llllll}
			\multicolumn{1}{c|}{} & \multicolumn{1}{c}{Name}  & \multicolumn{1}{c}{Symbol}  \\
			\hline
			\multirow{6}*{Computational quantities} & resting point of accommodation &  $A$    \\
			& resting point of convergence & $V$     \\
			& axial length of the eye &  $AL$  \\
			& the accommodative response &  $AR$  \\
			& the convergence response &  $VR$  \\
			& the trend of myopia &  $O$  \\
			\hline
			\multirow{7}*{Statistical 
				quantities
			} & the refractive error of the eye &  $M(diopter)$  \\
			& viewing distance &  $d(m)$  \\
			& the size of the pupil &  $P(mm)$  \\
			& type of near work being performed &  $n$  \\
			& the duration of near work &  $t(min)$  \\
			& the number of aberrations &  $W$  \\
			& the level of ambient lighting &  $L(lux)$  \\
			
	\end{tabular}}
	\label{table-params}
\end{table}

\begin{table}[!tbp]
	\center
	\caption{Experimental Initialization Values}
	\resizebox{0.4\linewidth}{!}{
		\begin{tabular}{l|llllll}
			\toprule
			\multicolumn{1}{c|}{age} & \multicolumn{1}{c}{$AL_0$}  & \multicolumn{1}{c}{$P_0$} & $M_0$ & $W_0$  \\
			\hline
			8-10 & 20 &  6  & 55  & 48   \\
			\hline
			10-12& 20 & 5 &   61   &   52     \\
			\hline
			12-14 & 22 &  3 & 68     & 51 \\
			\hline
			14-16 & 23 &  5 & 52     & 47 
	\end{tabular}}
	\label{table-ablation_1}
\end{table}

\begin{table}[!tbp]
	\center
	\caption{Experimental data for each age group}
	\resizebox{0.85\linewidth}{!}{
		\begin{tabular}{l | c | cc|cccc|ccc | l}
			\toprule
			\multirow{2}*{age} &  \multirow{2}*{n} & \multicolumn{2}{c}{\textbf{EAS}} & \multicolumn{4}{c}{\textbf{ECAS}} & \multicolumn{3}{c|}{\textbf{EVL}} &  \multirow{2}*{SER}  \\
			\cline{3-11}
			& & t & L &   P  & M & d  & W & AR & VR & O &   \\
			%				\midrule
			\hline
			8-10  & 1   & 30   & 987 & 8  & 55.01  & 0.1   & 53   & 106.18   & 117.196 & 0.906 &  -0.65 \\
			8-10  & 1   & 120  & 987 & 8  & 55.01  & 0.1   & 53   & 216.175  & 245.63  & 0.88  &  -1.1 \\
			10-12 & 1.5 & 30  & 987  & 6  & 61.0   & 0.1   & 54   & 111.572  & 123.773 & 0.904 & -0.83  \\
			10-12 & 1.5 & 30  & 207  & 6  & 61.0   & 0.1   & 54   & 154.32   & 179.44  & 0.86  & -1.5 \\
			
			12-14  & 2 & 30  & 987   & 4  & 68.01   & 0.1   & 55  & 132.096 & 145.19  & 0.91  & -1.2   \\
			12-14  & 2 & 30  & 987   & 4  & 68.01   & 0.05  & 55  & 130.818 & 150.365 & 0.87  & -2.4  \\
			14-16  & 1 & 30  & 987   & 8  & 52.101  & 0.1   & 54  & 94.525  & 99.727  & 0.947 & -1.1 \\
			14-16  & 1 & 60  & 207   & 8  & 52.101  & 0.1   & 54  & 205.127 & 250.154 & 0.82  & -3.2 
	\end{tabular}}
	\label{table-ablation_2}
\end{table}

\section*{Conflict of Interest Statement}
%All financial, commercial or other relationships that might be perceived by the academic community as representing a potential conflict of interest must be disclosed. If no such relationship exists, authors will be asked to confirm the following statement: 

The authors declare that the research was conducted in the absence of any commercial or financial relationships that could be construed as a potential conflict of interest.

\section*{Author Contributions}

XY contributed to conception and design of the study. GH organized the database. CF performed the statistical analysis. XJ wrote the first draft of the manuscript. XY wrote sections of the manuscript. All authors contributed to manuscript revision, read, and approved the submitted version.

\bibliographystyle{Frontiers-Harvard} %  Many Frontiers journals use the Harvard referencing system (Author-date), to find the style and resources for the journal you are submitting to: https://zendesk.frontiersin.org/hc/en-us/articles/360017860337-Frontiers-Reference-Styles-by-Journal. For Humanities and Social Sciences articles please include page numbers in the in-text citations 
\bibliography{test}

\begin{thebibliography}{16}
\providecommand{\natexlab}[1]{#1}
\expandafter\ifx\csname urlstyle\endcsname\relax
  \providecommand{\doi}[1]{doi:\discretionary{}{}{}#1}\else
  \providecommand{\doi}{doi:\discretionary{}{}{}\begingroup
  \urlstyle{rm}\Url}\fi
\providecommand{\selectlanguage}[1]{\relax}
\providecommand{\bibAnnoteFile}[1]{%
  \IfFileExists{#1}{\begin{quotation}\noindent\textsc{Key:} #1\\
  \textsc{Annotation:}\ \input{#1}\end{quotation}}{}}
\providecommand{\bibAnnote}[2]{%
  \begin{quotation}\noindent\textsc{Key:} #1\\
  \textsc{Annotation:}\ #2\end{quotation}}

\bibitem[{Bach et~al.(2019)Bach, Villegas, Gold, Shi, and
  Murray}]{bach2019axial}
Bach, A., Villegas, V.~M., Gold, A.~S., Shi, W., and Murray, T.~G. (2019).
\newblock Axial length development in children.
\newblock \emph{International journal of ophthalmology} 12, 815
\bibAnnoteFile{bach2019axial}

\bibitem[{Borsting et~al.(2003)Borsting, Rouse, Deland, Hovett, Kimura, Park
  et~al.}]{borsting2003association}
Borsting, E., Rouse, M.~W., Deland, P.~N., Hovett, S., Kimura, D., Park, M.,
  et~al. (2003).
\newblock Association of symptoms and convergence and accommodative
  insufficiency in school-age children.
\newblock \emph{Optometry (St. Louis, Mo.)} 74, 25--34
\bibAnnoteFile{borsting2003association}

\bibitem[{Grosvenor(2007)}]{grosvenor2007primary}
Grosvenor, T.~P. (2007).
\newblock \emph{Primary care optometry} (Elsevier Health Sciences)
\bibAnnoteFile{grosvenor2007primary}

\bibitem[{Han et~al.(2022)Han, Liu, Chen, and He}]{han2022myopia}
Han, X., Liu, C., Chen, Y., and He, M. (2022).
\newblock Myopia prediction: a systematic review.
\newblock \emph{Eye} 36, 921--929
\bibAnnoteFile{han2022myopia}

\bibitem[{Huang et~al.(2020)Huang, Hsiao, Tsai, Tsai, Chen, Hsu
  et~al.}]{huang2020protective}
Huang, P.-C., Hsiao, Y.-C., Tsai, C.-Y., Tsai, D.-C., Chen, C.-W., Hsu, C.-C.,
  et~al. (2020).
\newblock Protective behaviours of near work and time outdoors in myopia
  prevalence and progression in myopic children: a 2-year prospective
  population study.
\newblock \emph{British Journal of Ophthalmology} 104, 956--961
\bibAnnoteFile{huang2020protective}

\bibitem[{Jiang et~al.(2002)Jiang, Hung, and Ciuffreda}]{jiang2002models}
Jiang, B.-c., Hung, G.~K., and Ciuffreda, K.~J. (2002).
\newblock Models of vergence and accommodation-vergence interactions.
\newblock \emph{Models of the visual system} , 341--384
\bibAnnoteFile{jiang2002models}

\bibitem[{Kasthurirangan and Glasser(2006)}]{KASTHURIRANGAN20061393}
Kasthurirangan, S. and Glasser, A. (2006).
\newblock Age related changes in the characteristics of the near pupil
  response.
\newblock \emph{Vision Research} 46, 1393--1403.
\newblock \doi{https://doi.org/10.1016/j.visres.2005.07.004}
\bibAnnoteFile{KASTHURIRANGAN20061393}

\bibitem[{Li et~al.(2016)Li, Luo, Li, Zheng, Ji, Ma et~al.}]{li2016effects}
Li, B., Luo, X., Li, T., Zheng, C., Ji, S., Ma, Y., et~al. (2016).
\newblock Effects of constant flickering light on refractive status, 5-ht and
  5-ht2a receptor in guinea pigs.
\newblock \emph{PloS one} 11, e0167902
\bibAnnoteFile{li2016effects}

\bibitem[{Li et~al.(2015)Li, Li, Kang, Zhou, Liu, Li et~al.}]{li2015near}
Li, S.-M., Li, S.-Y., Kang, M.-T., Zhou, Y., Liu, L.-R., Li, H., et~al. (2015).
\newblock Near work related parameters and myopia in chinese children: the
  anyang childhood eye study.
\newblock \emph{PloS one} 10, e0134514
\bibAnnoteFile{li2015near}

\bibitem[{Matsuo and Ohtsuki(1992)}]{matsuo1992follow}
Matsuo, T. and Ohtsuki, H. (1992).
\newblock Follow-up results of a combination of accommodation and convergence
  insufficiency in school-age children and adolescents.
\newblock \emph{Graefe's archive for clinical and experimental ophthalmology}
  230, 166--170
\bibAnnoteFile{matsuo1992follow}

\bibitem[{{\"O}ner et~al.(2016){\"O}ner, Bulut, Oru{\c{c}}, and
  {\"O}zg{\"u}r}]{oner2016influence}
{\"O}ner, V., Bulut, A., Oru{\c{c}}, Y., and {\"O}zg{\"u}r, G. (2016).
\newblock Influence of indoor and outdoor activities on progression of myopia
  during puberty.
\newblock \emph{International ophthalmology} 36, 121--125
\bibAnnoteFile{oner2016influence}

\bibitem[{Pan et~al.(2018)Pan, Wu, Liu, Li, and Zhong}]{pan2018types}
Pan, C.-W., Wu, R.-K., Liu, H., Li, J., and Zhong, H. (2018).
\newblock Types of lamp for homework and myopia among chinese school-aged
  children.
\newblock \emph{Ophthalmic epidemiology} 25, 250--256
\bibAnnoteFile{pan2018types}

\bibitem[{Singh et~al.(2019)Singh, James, Yadav, Kumar, Asthana, and
  Labani}]{singh2019prevalence}
Singh, N.~K., James, R.~M., Yadav, A., Kumar, R., Asthana, S., and Labani, S.
  (2019).
\newblock Prevalence of myopia and associated risk factors in schoolchildren in
  north india.
\newblock \emph{Optometry and Vision Science} 96, 200--205
\bibAnnoteFile{singh2019prevalence}

\bibitem[{Smith et~al.(2012)Smith, Hung, and Huang}]{smith2012protective}
Smith, E.~L., Hung, L.-F., and Huang, J. (2012).
\newblock Protective effects of high ambient lighting on the development of
  form-deprivation myopia in rhesus monkeys.
\newblock \emph{Investigative ophthalmology \& visual science} 53, 421--428
\bibAnnoteFile{smith2012protective}

\bibitem[{Vienne et~al.(2014)Vienne, Sorin, Blond{\'e}, Huynh-Thu, and
  Mamassian}]{vienne2014effect}
Vienne, C., Sorin, L., Blond{\'e}, L., Huynh-Thu, Q., and Mamassian, P. (2014).
\newblock Effect of the accommodation-vergence conflict on vergence eye
  movements.
\newblock \emph{Vision research} 100, 124--133
\bibAnnoteFile{vienne2014effect}

\bibitem[{Wang et~al.(2023)Wang, Liu, Yang, Yu, Tong, Tang
  et~al.}]{wang2023short}
Wang, G., Liu, Y.-F., Yang, Z., Yu, C.-X., Tong, Q., Tang, Y.-L., et~al.
  (2023).
\newblock Short-term acute bright light exposure induces a prolonged anxiogenic
  effect in mice via a retinal iprgc-cea circuit.
\newblock \emph{Science Advances} 9, eadf4651
\bibAnnoteFile{wang2023short}

\end{thebibliography}

%%% Make sure to upload the bib file along with the tex file and PDF
%%% Please see the test.bib file for some examples of references

%%% If you don't add the figures in the LaTeX files, please upload them when submitting the article.
%%% Frontiers will add the figures at the end of the provisional pdf automatically
%%% The use of LaTeX coding to draw Diagrams/Figures/Structures should be avoided. They should be external callouts including graphics.

\end{document}